\documentclass[aps,prd,reprint,superscriptaddress,nofootinbib,showkeys]{revtex4-2}

\usepackage{amsmath,amssymb}
\usepackage{mathrsfs}
\usepackage{braket}
\usepackage{hyperref}
\usepackage{orcidlink}
\hypersetup{
  colorlinks=true,
  linkcolor=black,
  citecolor=black,
  urlcolor=black
}

\newcommand{\BQ}{\mathcal{B}_{Q}}
\newcommand{\rBQ}{r_{\rm qb}}
\newcommand{\nBQ}{n_{\rm qb}}
\newcommand{\lP}{\ell_{P}}
\newcommand{\Tqb}{\mathcal{T}_{\rm qb}}

\begin{document}

\title{The Quantum Boundary of Black Hole Interiors:
Signature Failure and the Termination of Mass Inflation}
 
\author{Edward J. Shaya\, \orcidlink{0000-0002-3234-8699}}
\email{eshaya2@gmail.com}
\affiliation{Department of Astronomy, University of Maryland, College Park, Maryland 20742, USA}

\date{\today}

\begin{abstract}
Classical general relativity predicts a singularity at the center of every black hole. We argue that it is never reached. We adopt the conjecture of \citet{shaya2026signature} that a metric element of non-Lorentzian signature is inadmissible: it cannot enter the Feynman sum over geometries, and so cannot belong to the physical manifold. Nothing beyond standard quantum mechanics is invoked. Signature is the only pointwise diffeomorphism-invariant content of the metric, so the criterion is local and coordinate independent. Transverse-traceless graviton vacuum fluctuations grow as $\Delta_g = \sqrt{8/\pi}\,\ell_P/q$ on a scale $q$, and any finite-scale metric stress driving an eigenvalue through zero removes that element. When the surviving elements can no longer sustain connected, differentiable, causally propagating support, the manifold terminates on a quantum boundary $\mathcal{B}_Q$ at finite radius. For the vacuum Schwarzschild interior the tidal Weyl stress gives $r_{\rm qb} \simeq 1.0\times10^{-22}(M/M_\odot)^{1/3}$~m for the adopted curvature response $C_W=0.15$. For a spinning hole the boundary forms much farther out: mass inflation at the inner horizon $r_{-}$ concentrates an exponentially growing interior mass into a thin layer whose thickness is set by the accretion rate and floored at one spatial quantum, and the counter-streaming null fluxes supply a trace-free Ricci stress along the radial axis alone. This caps the factor $n_{\rm qb}$ by which mass inflation can amplify the interior mass function, at between $10^{38}$ and $10^{69}$ for a $10\,M_\odot$ hole depending on its accretion history, and leaves the Cauchy horizon, the ring, and all deeper extensions outside the physical manifold. The Gibbons--Hawking--York term over this terminal slice yields a finite interior action, $S_{GHY}^{qb} \approx \frac{3}{2} n_{\rm qb} Mc^2\,\Delta t$.
\end{abstract}

\keywords{quantum gravity, black hole singularities, metric signature, sum over geometries, functional integral, mass inflation, ADM formalism, effective field theory of gravity}

\maketitle

\section{Introduction} \label{sec:introduction}

We present a framework for quantum gravity that may be ultraviolet finite within its domain without requiring trans-Planckian extension. The framework rests on the Wheeler--DeWitt (WDW) equation taken seriously as a non-perturbative statement about the quantum state of geometry, together with the Feynman sum over geometries. This approach was first articulated by Misner \citep{misner1957}, who established that in quantizing general relativity via the Feynman functional integral, the manifold points ``have no physical significance in themselves'' and introduced a sample state functional that vanishes identically for geometrically inadmissible metric configurations. The step taken here is to make Misner's inadmissibility concrete and local: a metric element belongs to the physical manifold only if it is nondegenerate and Lorentzian, and quantum fluctuations at small scales drive a growing fraction of elements out of that class.

The quantum boundary $\BQ$ is where the admissible domain of the gravitational functional integral terminates. The chain begins with a single element,
\begin{equation}
\hbox{signature failure}\;\Longrightarrow\;\hbox{no metric element} ,
\label{eq:admissibility_chain}
\end{equation}
which excises one element, a quantum topological hole (QTH) in the terminology of Sec.~\ref{subsec:signature}, and leaves the surrounding geometry intact. A boundary requires more than that. Differentiability is not a property of a single element: it requires a \emph{patch}, the minimum contiguous region of admissible elements on which derivatives of the metric are defined, which for a spatial quantum $q$ is a neighborhood of extent $3q$ in each direction. As failures proliferate, complete patches become rare and those that remain no longer form a connected network. It is the loss of connected differentiable support, not the loss of any individual element, that terminates the manifold:
\begin{equation}
\hbox{loss of patch connectivity}\;\Longrightarrow\;\Psi=0 .
\label{eq:connectivity_chain}
\end{equation}
Here signature failure does not mean that present calculations are incomplete, nor that some averaged geometry has been coarse-grained away. Lorentzian signature is what makes a metric a spacetime metric: it is what supplies light cones, a time orientation, a causal order, and a hyperbolic evolution problem \citep{ChoquetBruhat1952,ChoquetBruhatGeroch1969,Geroch1970,hawkingellis1973,Wald1984}. Where a quantum realization of $g_{\mu\nu}$ is not Lorentzian, the principal part of the field equations is ultrahyperbolic and codimension-one data no longer define the usual Cauchy problem \citep{Asgeirsson1937,John1938,CraigWeinstein2009}; quantum field theory on that element is equally undefined \citep{Dimock1980,krw1997,Wald1994}. Such a region is treated as absent from the physical spacetime, and configurations requiring its inclusion are assigned vanishing support in the effective geometrical wavefunctional.

The criterion is attractive precisely because it is not a postulate about what we can know. At a point one may always set $g_{\mu\nu}=\eta_{\mu\nu}$ and $\partial_\alpha g_{\mu\nu}=0$, so signature is the \emph{only} pointwise diffeomorphism-invariant content the metric carries. Whether a given configuration in the sum over geometries is Lorentzian at a given point is a property of that configuration, not of the observer, the slicing, or the approximation scheme.

Several prior approaches have sought to resolve the UV problem of quantum gravity by extending or replacing the theory at the Planck scale. For instance, asymptotic safety postulates a non-trivial UV fixed point to render the theory renormalizable at all trans-Planckian energies \citep{weinberg1979,reuter1998}. Loop quantum cosmology and Planck star proposals utilize discrete area spectra to replace the continuous metric manifold with fundamentally discrete structures (e.g., spin networks or simplicial complexes), kinematically forcing a quantum bounce \citep{bojowald2001,ashtekar2018,rovelli2014}. In each of these cases, fundamentally new trans-Planckian physics or novel degrees of freedom must be injected at the Planck scale. In contrast, the $\BQ$ framework operates on parsimony. Rather than extending the physics, it finds that the continuous metric manifold and the domain of the Feynman--DeWitt measure contain strict limits to their own definability, and that those limits are already visible in the ordinary quantized graviton field.

The argument proceeds as follows. We first set up the WDW framework with an enlarged configuration space that admits locations carrying no metric element, and state the signature criterion that decides membership. We then summarize the quantitative results of the companion study \citep{shaya2026signature}: the vacuum fluctuation amplitude at scale $q$, the minimum spatial quantum $q_{\min}\simeq1.9\,\lP$ below which connected differentiable support cannot be sustained, and the thresholds in curvature-induced metric stress at which that support is lost. We examine black holes as physical realizations of $\BQ$, treating the vacuum tidal case in closed form and then the mass-inflation case, where the relevant stress is the trace-free Ricci stress of counter-streaming null fluxes rather than the tidal Weyl field. This yields a quantitative cutoff for the mass-inflation instability and a Lorentzian evaluation of the interior boundary action. Four accretion histories, from the isolated Kerr hole to mergers, bracket how deep the boundary lies in realistic cases. We then take up two consequences: the suppression of BKL oscillation, since the chaotic approach to a singularity is cut off before it can develop, and the status of weak and strong cosmic censorship once the Cauchy horizon lies outside the physical manifold. 

\section{The Wheeler--DeWitt Framework and Metric Admissibility}
\label{sec:wdw-framework}

The canonical quantization of general relativity leads, via the Dirac procedure applied to the Hamiltonian and diffeomorphism constraints, to the Wheeler--DeWitt equation \citep{dewitt1967}:
\begin{equation}
	\hat{H}\,\Psi[h_{ij}] = 0, \label{eq:WDW}
\end{equation}
where $\Psi[h_{ij}]$ is the wavefunctional on superspace, the space of Riemannian 3-metrics $h_{ij}$ on a spatial hypersurface $\Sigma$, modulo diffeomorphisms, and $\hat{H}$ is the quantized scalar constraint. 
This is supplemented by the momentum constraints $\hat{H}^i\Psi = 0$ to ensure the wavefunctional is defined on the diffeomorphism-invariant configuration space (superspace).

The complex distribution $\Psi$ on the boundary of spacetime is determined by the Lorentzian form of the Feynman sum over geometries (the gravitational functional integral) \citep{misner1957,hartlehawking1983}:
\begin{equation}
	\Psi[h_{ij}] = \mathcal{N}\int_{g|_{\partial\mathcal{M}}=h_{ij}}
\!\!\mathcal{D}[g_{\mu\nu},\varnothing]\,\mathcal{D}[\varphi]\;
	e^{iS[g,\varphi]/\hbar},
\label{eq:functionalintegral}
\end{equation}
where $\mathcal{D}[g_{\mu\nu},\varnothing]$ denotes the formal, gauge-fixed integration over diffeomorphism-inequivalent four-geometries \emph{together with} a sum over which locations carry no admissible metric element at all, $\mathcal{N}$ is a normalizing factor, $h_{ij}$ denotes the induced metric on the \emph{full} boundary $\partial\mathcal{M}$ of the four-dimensional manifold $\mathcal{M}$, and $S[g,\varphi] = S_\mathrm{EH}[g] + S_\mathrm{matter}[g,\varphi]$ is the total action, including the Einstein--Hilbert action:

\begin{equation}
  S_\mathrm{EH}[g] = \frac{c^3}{16\pi G} \int_\mathcal{M}\!d^4x\,\sqrt{-g}\,(R - 2\Lambda) + S_\mathrm{GHY},
\end{equation}
where $S_\mathrm{GHY}$ is the Gibbons--Hawking--York boundary term \citep{gibbonshawking1977} integrated over all components of $\partial\mathcal{M}$. The symbol $\varnothing$ is not an additional field value and $\mathcal{D}[g,\varnothing]$ is not a new fundamental measure: it is shorthand for the enlarged configuration space obtained by summing over manifolds with different excision content, developed in \citep{shaya2026signature}. Specifying which locations are excised is itself part of specifying the manifold, so the domain of integration is generated rather than presupposed, and the edge of the excised set is a genuine boundary of the domain. That is where the GHY term enters, and it is why every excised region, however small, carries its own boundary contribution. In the present framework the inner $\BQ$ surface is the last admissible boundary of the continuous geometry. Boundary data are therefore taken on the admissible side of $\BQ$; configurations that require a continuation into the inadmissible region are assigned the terminal condition $\Psi=0$.

The Feynman sum inherently evaluates the probability amplitude by weighting all possible spacetime histories by the phase factor $e^{iS/\hbar}$.
In the stationary phase approximation, the functional integral is dominated by the classical action $S_{cl}$, where the variation of the action is zero ($\delta S = 0$).
Because the Wheeler--DeWitt framework lacks an external time parameter, this saddle-point solution represents the entirety of a static 4D block universe rather than a dynamically evolving state.
Quantum fluctuations around this block solution are permitted within a variance bounded by variations of order $\delta S \sim \hbar$.
Through the canonical Arnowitt--Deser--Misner (ADM) formalism, the spatial metric $h_{ij}$ and its conjugate momentum $p^{ij}$ are fundamentally linked to the action via the Hamilton--Jacobi relation. 
This manifests as the canonical commutation relation of the WDW framework:

\begin{equation}
	\bigl[h_{ij}(x),\,p^{kl}(x')\bigr] = i\hbar\,\delta^{(k}_i\delta^{l)}_j\,\delta^{(3)}(x-x').
	\label{eq:CCR}
\end{equation}

Integrating this relation over a characteristic physical correlation volume $V$ yields the Heisenberg uncertainty principle for the gravitational field  $\langle \Delta p \rangle \langle \Delta h \rangle \sim \frac{\hbar}{V}$. It is this relation, evaluated at small $V$, that produces the fluctuation amplitude of Sec.~\ref{subsec:vacuum-amplitude} and drives the signature failures that follow.

\subsection{Signature as the Admissibility Criterion}
\label{subsec:signature}

A metric element of physical size $q$ is called \emph{admissible} when its local $g_{\mu\nu}$ is nondegenerate and has Lorentzian signature $(-,+,+,+)$, and inadmissible otherwise. When a quantum realization is not Lorentzian, that elementary metric is absent from the physical manifold, and the corresponding four-dimensional excision is a \emph{quantum topological hole} (QTH) \citep{shaya2026signature}. An isolated QTH is surrounded topologically by an $S^3$ and alters the Betti-number content of the manifold.

Three consequences follow from failure, and together they are why no weaker reading is available. First, if more than one direction becomes timelike, the timelike directions at the point form a single connected set, so there is no two-sheeted cone and no time orientation to be chosen: causal order does not exist there \citep{HawkingKingMcCarthy1976,Malament1977}. Second, the field equations become ultrahyperbolic, and the Cauchy problem from ordinary codimension-one data is not well posed \citep{Asgeirsson1937,John1938,CraigWeinstein2009}. Third, no hypersurface through the point carries a Riemannian induced metric; this is linear algebra rather than a statement about any particular slicing, and it is what makes the criterion foliation independent. The damage to the ADM formulation follows from the third: superspace is built on positive-definite three-metrics, and at such a point it has no admissible element.

The criterion should not be described as the requirement of well-posedness as such, since a Riemannian $(+,+,+,+)$ element is also well posed. What it selects is the single signature on which the field equations describe evolution from data.

A boundary built from this criterion is not a wall standing in space. Wherever it forms in a black hole interior it lies within a horizon, in a region where the radial coordinate $r$ is timelike, so a surface of constant $r$ is spacelike. It is a last moment rather than a place, reached by every part of an infalling body at once; no part of one could be extended through it while the rest remained behind, any more than a hand could be put through the final moment of a collapsing universe. This is what distinguishes the excision here from the pathologies of a region cut out of space, which would be bounded by a timelike surface.

It is essential that the test be applied to the four-metric rather than to a three-metric on a chosen slice. Painlev\'e--Gullstrand slicing \citep{Painleve1921,Gullstrand1922} keeps $h_{ij}=\delta_{ij}$ through a horizon and to arbitrarily large curvature, so a criterion applied to $h_{ij}$ alone reports no failure inside a black hole even where none is warranted. The four-metric criterion has no such false negative. It also does not disqualify an ordinary black hole interior merely because familiar coordinates exchange character across a horizon: the metric remains admissible wherever the four-metric remains nondegenerate and Lorentzian, which inside the horizon it does.

The connection between Eq.~\eqref{eq:functionalintegral} and the signature calculation is direct, and requires no additional measurement postulate. Signature is configuration diagonal: for a metric-configuration eigenstate, whether the local four-metric is Lorentzian is an algebraic property of that same configuration. If $\mathcal{L}$ denotes the Lorentzian region of metric-configuration space, the Born probability on an equal-time slice is
\begin{equation}
 P_{\rm L}
 =\frac{\displaystyle\int_{\cal L}\mathcal{D}[\delta g]\,|\Psi[\delta g]|^2}
       {\displaystyle\int\mathcal{D}[\delta g]\,|\Psi[\delta g]|^2}.
\label{eq:bornsignature}
\end{equation}
For the free graviton vacuum the equal-time wavefunctional is a real positive Gaussian functional of the transverse-traceless (TT) mode amplitudes, so sampling that Gaussian is precisely sampling $|\Psi_0[\delta g]|^2$, and the Lorentzian classification is applied configuration by configuration.

It might be objected that a metric always finds a way to reorganize itself
rather than fail. Horizon formation is the familiar example: at $r_s$ the
entries $g_{tt}$ and $g_{rr}$ exchange signs together, so that $r$ becomes the
timelike direction and the signature is preserved. If reaching the
Schwarzschild mass automatically produces that rearrangement, one might expect
a fluctuation of comparable size to be able to do the same. It cannot, for two
reasons. First, the horizon exchange is not a signature change at all: in
Kruskal coordinates the metric is regular at $r_s$ and its eigenvalue count is
unchanged there, the apparent swap being a property of the Schwarzschild chart.
The criterion here is applied to the eigenvalues of $g_{\mu\nu}$ in a local
orthonormal frame, so it is blind to that rearrangement, and equally a chart in
which entries appear to swap cannot restore a count that has genuinely changed.
Second, a horizon arises as a solution of the constraint equations sourced by
stress-energy and is organized to match that source, whereas a
transverse-traceless vacuum fluctuation carries no source: its renormalized
stress-energy is subtracted by construction, and there is nothing in a
Planck-sized region for the field equations to respond to. Nothing drives the
fluctuating metric toward the Schwarzschild pattern. What does remain open is
that at $\Delta_g \sim 1$ the free-field description of the fluctuation is
itself at the edge of its validity, so the behavior of the fully nonlinear
vacuum at that amplitude is not settled by the linear calculation.

Physical spacetime in the $\BQ$ framework is therefore a manifold with boundary: beyond the boundary the description assigns no space. This is consistent with the standard treatment of singularities in general relativity, in which Hawking and Ellis \citep{hawkingellis1973} define singular points as having been ``cut out of space-time,'' with geodesic incompleteness \citep{penrose1965} as the signal that this has occurred. What the present criterion supplies is a candidate quantum-geometric mechanism that identifies where and why the cut occurs at $\rBQ > 0$, rather than at $r=0$ by fiat. Geodesics end not because they meet a point of infinite density but because there is no spacetime beyond $\BQ$ through which they could continue.

The zero-support condition itself is not new. DeWitt \citep{dewitt1967} proposed that the wavefunctional should vanish on singular three-geometries, $\Psi[^{(3)}\mathcal{G}] = 0$, a prescription now standard in quantum cosmology and widely used as a criterion for singularity resolution \citep{kiefer2019}. It has been applied to the black hole interior directly: solving the Wheeler--DeWitt equation on Kantowski--Sachs slices of the Schwarzschild interior yields a wavefunction that vanishes as $r \to 0$ \citep{brahmayeom2022,Paul_2025}, and \citep{brahmayeom2022} further argue that the BKL fracturing of the interior into causally disconnected sub-volumes forces vanishing boundary conditions on each. Two differences distinguish the present treatment. First, DeWitt's condition is imposed on solutions, and the black hole applications obtain $\Psi \to 0$ by solving the constraint and finding that the amplitude decays; here the vanishing is a statement about what exists rather than about decay, analogous to a particle in an infinite well, where $\psi$ vanishes identically outside not because the amplitude decays but because those configurations are excluded. Second, and consequently, the condition is not imposed at a singular three-geometry. It is imposed where admissibility fails, which occurs at finite volume and finite curvature, on a surface the classical solution reaches well before any singularity.

\subsection{The Vacuum Amplitude and the Minimum Spatial Quantum}
\label{subsec:vacuum-amplitude}

The quantitative inputs used below are established in the companion paper \citep{shaya2026signature} and are summarized here.

For tensor perturbations $\chi_{ij}$ about a spatially flat Friedmann background, quantized in the standard way \citep{Grishchuk1974,MFB1992}, the two-polarization vacuum spectrum is $P_T(k_{\rm phys}) = 2\hbar k_{\rm phys}^2/\pi^2 \overline M_P^2$ with $\overline M_P^2 = \hbar c/8\pi G$. Defining the cutoff length $q$ by the angular wavenumber $k_{{\rm phys},\max}=1/q$ and the rms fractional spatial-metric fluctuation by $\Delta_g^2 \equiv \langle \chi_{ij}\chi^{ij}\rangle$, integration of the spectrum gives
\begin{equation}
	\Delta_g(q) \;=\; \sqrt{\frac{8}{\pi}}\,\frac{\lP}{q}
	\;=\; \frac{1.596}{\alpha},
	\qquad q \equiv \alpha\,\lP .
	\label{eq:Dgq}
\end{equation}
All dependence on the scale factor cancels, so Eq.~\eqref{eq:Dgq} is a statement about physical length alone. This is the Planck-over-scale behavior anticipated in quantum geometrodynamics \citep{dewitt1967,Wheeler1968}, here with its coefficient fixed by the TT vacuum normalization and the stated angular cutoff, and nothing more.

Two features of Eq.~\eqref{eq:Dgq} control everything that follows. It is history free: no factor of $a$, no expansion rate, and no matter content appears, so the same amplitude applies in a cosmological collapse and in a black hole interior. And it grows without bound as $q$ falls, so signature failure becomes important at a length of order $\lP$ whether or not any classical curvature is present.

One-point statistics are obtained from the induced ensemble. After angular averaging, the TT covariance is
$\langle \chi_{ij}\chi^{kl}\rangle = (\Delta_g^2/10)(\delta_i^{\,k}\delta_j^{\,l}+\delta_i^{\,l}\delta_j^{\,k}-\tfrac{2}{3}\delta_{ij}\delta^{kl})$,
which is a rotationally invariant traceless Gaussian orthogonal ensemble \citep{Wigner1958,Dyson1962,Mehta2004}. Contracting on both pairs returns $\langle \chi_{ij}\chi^{ij}\rangle = \Delta_g^2$, which fixes the normalization and matches the definition of $\Delta_g$ through the four-metric contraction in \citep{shaya2026signature}. Since $\delta g_{0\mu}=0$ in TT gauge, Sylvester's law of inertia makes the four-metric Lorentzian if and only if $\delta_{ij}+\chi_{ij}$ is positive definite. Monte Carlo sampling and an independent semi-analytic eigenvalue integral agree that $P_{\rm L}$ falls from unity at $\Delta_g=0.4$ to $0.92$ at $\Delta_g=1.0$ and $0.48$ at $\Delta_g=1.6$ \citep{shaya2026signature}. Note the scale of the fluctuation required: at $\Delta_g\simeq0.85$, where $P_{\rm L}$ has begun to decline appreciably, the rms of one diagonal component is only about $0.31$. Signature fails long before any single metric component is of order unity, because it is the smallest \emph{eigenvalue} that must cross zero.

A one-point probability is not the boundary condition, however. A local classical geometry requires second derivatives, so the operative requirement is that an element sit at the center of a complete differentiable neighborhood, and that such neighborhoods form a connected set that persists causally. In \citep{shaya2026signature} this is implemented on a periodic lattice of spacing $q$: sites are classified Lorentzian or not from the correlated TT field, a site is in the differentiable state only if its full $3\times3\times3$ neighborhood is active, the active set is updated once per light-crossing time $q/c$ with one-cell causal repair from surviving support, and only differentiable components with nonzero periodic winding are retained. The result is a minimum spatial quantum
\begin{equation}
	q_{\min} \;\simeq\; 1.9\,\lP,
	\qquad
	\Delta_{g,c} \;\simeq\; 0.85 ,
	\label{eq:qmin}
\end{equation}
below which the vacuum alone destroys connected differentiable support. A gauge-reduced full four-metric ensemble gives $\simeq 2.0\,\lP$, so the restriction to the TT sector does not control the answer.

That a resolution limit of order $\lP$ should exist has long been argued on independent grounds \citep{Mead1964,garay1995,Hossenfelder2013}, and the combination $\sqrt{\lP L}$ appears in spacetime-foam estimates \citep{ngvandam1994}, there with $L$ a propagation distance. The familiar Compton--Schwarzschild estimate places the crossover where a virtual excitation confined to $q$ has $r_s(q)=2\lP^2/q$ equal to $q$, that is at $q=\sqrt{2}\,\lP$. Equation~\eqref{eq:qmin} lies \emph{above} this: at $q_{\min}$ one has $r_s/q < 2(\lP/q_{\min})^2 \simeq 0.55$. The manifold loses differentiable Lorentzian support before a spatial quantum becomes smaller than the region closed by its own localization energy, because the smallest eigenvalue of a fluctuating symmetric metric can cross zero while every individual component remains well below unity, whereas closure requires an order-unity excursion. Signature failure is the first thing to go wrong, not the last.

Equation~\eqref{eq:qmin} is a floor on the size of a spatial quantum, not yet a boundary in spacetime. In flat or weakly curved regions the manifold simply operates at $q > q_{\min}$ and nothing happens. What produces a boundary is metric stress. Across an element of size $q$ the curvature deforms the metric by an amount of order $|R|q^2$, and that deterministic deformation adds to the fluctuation, depressing one or more eigenvalues toward zero. The threshold is set by the deformation across the element, not by the magnitude of any curvature invariant at a point.

\subsection{Curvature-Induced Metric Stress}
\label{subsec:stress}

Write the local metric of one spatial quantum as
\begin{equation}
	H_{ij}(q,t) \;\equiv\; \frac{\gamma_{ij}(q,t)}{a^2(t)}
	\;=\; \delta_{ij} + s_{ij}(q,t) + \chi^{Q}_{ij}(q,t),
	\label{eq:stressmetric}
\end{equation}
where $\gamma_{ij}$ is the spatial metric of the element, $a(t)$ the common isotropic factor, $\chi^{Q}_{ij}$ the quantum TT fluctuation of Eq.~\eqref{eq:Dgq}, and $s_{ij}$ the deformation induced by classical or semiclassical curvature across an element of size $q$. Dividing by $a^2$ cannot change the signs of the spatial eigenvalues, so it does not affect the admissibility of the spatial block, and within that block it is the \emph{anisotropic} part of $s_{ij}$ that matters: it changes the shape of the local metric and can carry its smallest eigenvalue through zero. The scale factor is not thereby irrelevant to the four-metric. A finite element spans an interval in time as well as in space, and $\ddot a$ enters its time-time component, so a spatially homogeneous geometry still deforms in four dimensions even when $H_{ij}$ is undeformed.

The general curvature content is displayed by the irreducible decomposition of the Riemann tensor into the Weyl tensor $C_{\alpha\beta\gamma\delta}$, the trace-free Ricci tensor $S_{\mu\nu} \equiv R_{\mu\nu}-\tfrac14 g_{\mu\nu}R$, and the scalar curvature $R$. Only the first two are anisotropic within the spatial block, and the two limits that concern this paper isolate them cleanly: a Ricci-flat vacuum supplies pure Weyl stress, while counter-streaming null fluxes supply pure trace-free Ricci stress with $R=0$. In a spatially homogeneous geometry the Weyl tensor vanishes identically, so any deformation comes from the Ricci sector alone: from the trace-free part when the content is radiation, which has $T^\mu{}_\mu = 0$ and hence $R=0$ for negligible cosmological constant, and from $R$ only for content of nonzero trace. That is the regime of a universe expanding from or contracting to Planck density \citep{shaya2026_mwi,shaya2026signature}. The patterns scanned here have $R=0$, apart from the matter sheet of Sec.~\ref{sec:black-holes}.

For the Weyl case, the kinematic manifestation of the tidal field along a timelike congruence is shear. For a congruence of four-velocity $u^\mu$, with $h_{\mu\nu} \equiv g_{\mu\nu} + u_\mu u_\nu/c^2$ the projector onto the space orthogonal to $u^\mu$, the gradient decomposes as
\begin{equation}
	\nabla_\nu u_\mu = \tfrac13\theta h_{\mu\nu} + \sigma_{\mu\nu}
	+ \omega_{\mu\nu} - u_\mu a_\nu ,
	\label{eq:kinematic}
\end{equation}
with $\theta = \nabla_\mu u^\mu$ the volume expansion, $\omega_{\mu\nu}$ the vorticity, and $a_\mu = u^\nu\nabla_\nu u_\mu$ the four-acceleration. Only the shear $\sigma_{\mu\nu}$, the symmetric trace-free part, changes the shape of an element at fixed volume. With the natural time over which one quantum deforms being its own light-crossing time $q/c$,
\begin{equation}
	s_{ij}(q,t) \;\sim\; \frac{q}{c}\,\sigma_{ij}(q,t) .
	\label{eq:sheartos}
\end{equation}
The electric Weyl tensor relative to the congruence, $E_{\mu\nu}\equiv C_{\mu\alpha\nu\beta}u^\alpha u^\beta/c^2$, is spatial, symmetric, and trace free, and gives the dimensionless finite-$q$ deformation
\begin{equation}
	\mathcal{T}^{(W)}_{ij}(q) \;\equiv\; -q^2 E_{\hat i \hat j},
	\label{eq:tidalstrength}
\end{equation}
and, for the Ricci case,
\begin{equation}
	s^{(R)}_{\hat\mu\hat\nu} \;=\; -C_R\,q^2\,S_{\hat\mu\hat\nu} ,
	\label{eq:ricciresponse}
\end{equation}
the sign in each following the geodesic-deviation acceleration
$D^2\xi^i/D\tau^2 = -c^2E^i{}_j\xi^j$. For a fixed eigenvalue pattern we write
$\mathcal{T}^{(W)}_{ij}=\mathcal{T}Q_{ij}$, and denote by $\Tqb(\alpha)$ the
threshold, for that pattern, in the metric deformation $C_W\mathcal{T}$ or
$C_R\mathcal{T}$ at which connected differentiable support is lost; these are obtained from ensemble scans and are
evaluated case by case in Sec.~\ref{sec:black-holes}. The coefficients
$C_W$ and $C_R$ convert curvature to metric deformation at finite $q$;
Appendices~\ref{app:CR} and \ref{app:CW} evaluate them from the
normal-coordinate expansion, giving $C_W \simeq C_R \simeq 0.15$; every
numerical value below is quoted for that choice, the symbolic forms retaining
the coefficients.

A general statement follows immediately and applies to every case treated in Sec.~\ref{sec:black-holes}. At fixed $q$, any evolution for which $(q/c)(\sigma_{\mu\nu}\sigma^{\mu\nu})^{1/2}\to\infty$, or for which the tidal deformation on scale $q$ grows without bound, must cross the finite threshold. The classical black hole interior has both properties. The boundary is therefore not a feature that must be arranged; it is unavoidable once the criterion is granted, and the only question is where it occurs.

\subsection{Causal Inadmissibility and the Kay--Radzikowski--Wald Theorems}
\label{subsec:causal-inadmissibility}

The matter sector:
\begin{equation} 
 \mathcal{Z}_{\mathrm{matter}}[g] = \int\mathcal{D}[\varphi]\;
	e^{iS_{\mathrm{matter}}[g,\varphi]/\hbar},
\label{eq:Z}
\end{equation}
is defined via the Feynman propagator, whose standard time ordering requires a globally hyperbolic spacetime. It must admit a smooth foliation by spacelike Cauchy surfaces. On spacetimes containing closed timelike curves or Cauchy horizon breakdowns, time-ordering is not globally definable, and the propagator is not a standard globally defined distribution.

The Kay--Radzikowski--Wald theorems \citep{krw1997} make this precise for a spacetime with a compactly generated Cauchy horizon, the case relevant to a retained CTC region. The two-point function of any Hadamard state defined on the initial globally hyperbolic region is necessarily singular at every base point of the horizon, so that renormalized quantities such as $\langle\varphi^2\rangle$ and the stress-energy tensor are ill defined there; no extension of the field algebra satisfies $F$-locality at such a point. Without such a regular state, $\mathcal{Z}_{\mathrm{matter}}[g]$ is not a standard semiclassical object. If an idealized geometry retained a traversable Cauchy horizon or closed timelike curve (CTC) region, the matter sector would fail as a functional integral contribution. The theorems assume a compactly generated horizon and so bear on that CTC region rather than on the inner horizon of a black hole, which is not compactly generated.

This is a weaker and more familiar failure than signature failure, and the two act in the same direction. Where signature fails there is no metric at all, hence no propagator to define; where signature survives but the causal structure is pathological, the matter sector is already ill defined on a perfectly good Lorentzian metric. As detailed in Sec.~\ref{subsec:case-kerr-inflation}, mass inflation closes the idealized loophole before the non-hyperbolic region becomes part of the physical interior, so neither failure is ever actually reached.

\section{Black Holes as Physical Realizations of $\BQ$} \label{sec:black-holes}

Every astrophysical black hole carries angular momentum, so we must consider
the Kerr black hole. The spherically symmetric solution is nonetheless
the natural starting point, since it is the case in which the boundary can be
located in closed form, and it fixes the length scale that the rotating cases
inherit.

\subsection{The Spherical Reference Radius}
\label{subsec:schwarzschild-radius}

The interior of a nonrotating Schwarzschild hole is Ricci flat, so no matter
stress is required and the deformation of Eq.~\eqref{eq:stressmetric} is
carried entirely by the tidal Weyl field. In a local orthonormal frame the
relative tidal acceleration across a separation $q$ is of order
$c^2(r_g/r^3)\operatorname{diag}(2,-1,-1)$ with $r_g \equiv GM/c^2$, so that
$E^{\rm Schw}_{\hat i\hat j} = -(r_g/r^3)\operatorname{diag}(2,-1,-1)$ and
Eq.~\eqref{eq:tidalstrength} gives
\begin{equation}
	\mathcal{T}^{(W)}_{{\rm Schw},ij} = \mathcal{T}\operatorname{diag}(2,-1,-1),
	\qquad
	\mathcal{T}(r,q) = \frac{r_g q^2}{r^3} .
	\label{eq:schwT}
\end{equation}
The local quantum metric is therefore
\begin{equation}
	\begin{split}
	H_{ij}(r,q,t) = {}&
	\operatorname{diag}\bigl(1+2C_W\mathcal{T},\,1-C_W\mathcal{T},\\
	&\qquad\quad 1-C_W\mathcal{T}\bigr)
	+ \chi^Q_{ij}(q,t).
	\end{split}
	\label{eq:schwarzschildH}
\end{equation}
This is a local finite-scale representation of the tidal deformation of one
spatial quantum, not an accumulated infall history. Note what does
\emph{not} happen: by Eq.~\eqref{eq:Dgq} the quantum amplitude $\Delta_g$ is
fixed by $q$ alone and does not grow as the radius falls. What grows is
$\mathcal{T}\propto r^{-3}$, which pushes the two transverse eigenvalues
down toward zero and thereby increases the fraction of realizations in which
one of them crosses. The boundary is where those failures are dense enough to
destroy connected differentiable support.

Scanning the tidal strength at fixed $\alpha$ with the same correlated TT
field and causal update described in Sec.~\ref{subsec:vacuum-amplitude} gives
the thresholds $\Tqb(\alpha)$ of Table~\ref{tab:tqb} \citep{shaya2026signature},
and hence
\begin{equation} 
	\begin{split}
	\rBQ(\alpha) &= C_W^{1/3}
	\left[\frac{\alpha^2}{\Tqb(\alpha)}\right]^{1/3}
	\left(r_g\lP^2\right)^{1/3}\\
	&\equiv C_{\rm qb}\left(r_g\lP^2\right)^{1/3}.
	\end{split}
\label{eq:rBQ}
\end{equation}

\begin{table}[ht]
\centering
\small
\caption{Critical Schwarzschild tidal strength $\Tqb$ for loss of connected
differentiable support at fixed spatial quantum $q=\alpha\lP$, and the
resulting radius coefficient $C_{\rm qb}=[C_W\alpha^2/\Tqb]^{1/3}$, evaluated at the adopted $C_W=0.15$.
Excerpted from \citep{shaya2026signature}.}
\begin{tabular*}{\linewidth}{@{\extracolsep{\fill}}|lll|@{}}
	\hline
	$\alpha$ & $\Tqb$ & $C_{\rm qb}$ \\
	\hline
	1.9 & 0.06 & 2.04 \\
	2.2 & 0.25 & 1.43 \\
	2.5 & 0.35 & 1.39 \\
	3.0 & 0.46 & 1.43 \\
	4.0 & 0.60 & 1.58 \\
	10.0 & 0.84 & 2.61 \\
	\hline
\end{tabular*}
\label{tab:tqb}
\end{table}

\noindent
The manifold survives at radius $r$ so long as \emph{some} admissible quantum
size does, so the physical boundary is the minimum of Eq.~\eqref{eq:rBQ} over
$\alpha$. That minimum is shallow and lies near $\alpha\simeq2.5$--$2.7$, with
$C_{\rm qb}\simeq1.39$. Quanta near the floor $q_{\min}$ of
Eq.~\eqref{eq:qmin} fail farther out because their own vacuum fluctuations
already sit near the fragmentation threshold; large quanta also fail farther
out because the tidal variation across a cell grows as $q^2$. The competition
between the two produces an intermediate scale that penetrates deepest. Thus
\begin{equation}
	\begin{split}
	\rBQ &\simeq 1.4\left(r_g\lP^2\right)^{1/3}\\
	&\simeq 1.0\times10^{-22}
	\left(\frac{M}{M_\odot}\right)^{1/3}\!\mathrm{m},
	\end{split}
\label{eq:rBQnum}
\end{equation}
for the adopted $C_W=0.15$ of Appendix~\ref{app:CW}, some thirteen orders of
magnitude above $\lP$. The interior spacetime
terminates at this quantum-geometric radius, and the pathological infinite
energy tensor at $r=0$ is absent from the manifold.

This spherical treatment is an idealization, and the spin needed to invalidate
it is remarkably small. Essentially every physical black hole possesses an
inner horizon lying far outside the spherical boundary radius, so
Eq.~\eqref{eq:rBQnum} is inapplicable to it. In Kerr the counter-streaming
instability at the inner horizon carries the metric stress to threshold at
$r \approx r_{-}$, far outside any radius at which the spherical tidal
criterion would act, and $\BQ$ forms there.

The boundary is self-stabilizing. Displaced outward, the metric
stress is below threshold, connected differentiable support survives, and the
functional integral has nonzero support ($\Psi \neq 0$), moving the boundary
back inward; displaced inward, the stress is above threshold and no connected
support is retained ($\Psi = 0$).

The spin itself evolves with accretion, and $\BQ$ follows. Since $r_{-} =
r_g(1-\epsilon)$ with $\epsilon = \sqrt{1-a^2/r_g^2}$, spin-up carries the
inner horizon outward and spin-down carries it inward: the boundary of the
manifold is a hypersurface $\rBQ(v)$ that tracks $r_{-}(v)$ at the small
offset of Eq.~\eqref{eq:dlnr}. No inconsistency arises when it moves inward,
because $\BQ$ is spacelike rather than a wall standing in space; each epoch
contributes its own terminal slice, not a persistent excision that a later one
would have to re-enter. The requirement is only that the hypersurface remain
spacelike, which fails if $r_{-}$ moves too quickly. That condition is
$|\mathrm{d}r_{-}/\mathrm{d}v| \ll 1$ in geometric units, and since the spin
changes on the accretion timescale the rate is of order the dimensionless
accretion rate $\mu$ of Sec.~\ref{subsec:mi-mechanism}, at most
$5\times10^{-7}$ in the cases considered here. The terminal surface therefore
stays firmly spacelike throughout.

Nor does the manifold grow back into the excised region, although in the
percolation description a failed element is not permanently lost: where the
local deformation is sub-threshold, failure is a fluctuation of the vacuum
ensemble and an element adjacent to admissible neighbors can be Lorentzian
again in the next realization. Inside $\BQ$ the failures are not stochastic.
The deterministic deformation already exceeds threshold, so the Lorentzian
fraction is driven toward zero and the isolated survivors cannot be assembled
into the $3q$ patch that differentiability requires; with no patch there are no
derivatives, hence no field equations to carry the geometry inward; and the
deformation increases monotonically inward, so an element restored at the edge
would fail again immediately.

Several further features of $\BQ$ are common to all four cases below and may
be stated once. At any advanced time the boundary is a single closed
two-surface bounding a region absent from the manifold, and because the manifold terminates at $\rBQ>0$ no causal
curve extends past it. It lies in every case inside a horizon, where $r$ is
timelike, so it is a spacelike surface: a terminal moment of the interior
rather than a wall that could be approached from one side and reached through. The Feynman sum accordingly runs over geometries that
terminate at this boundary rather than continuing through it. A geometry whose
causal curves are extended past $\rBQ$ is not admissible, not because such
curves are forbidden, but because their continuation would require metric
elements that do not exist: the extension is undefined rather than disallowed.
Nothing constrains the motion of matter within an admissible geometry, and an
infalling worldline reaches $\BQ$ in finite proper time and ends there. The
excision is not a manual surgery, but the geometrical representation of the
zero-support condition assigned where no Lorentzian metric element exists.

What distinguishes the four cases is only
where the threshold is first attained, which the spin and the accretion
history decide. We take them in order: the vacuum solution with no infall;
slow infall onto a rapidly spinning hole, the realistic astrophysical case;
slow infall onto a weakly spinning hole; and the high-accretion case
represented by mergers and by the assembly of the hole itself. Before the
cases, we treat the mechanism they share.

\subsection{Mass Inflation and the Extinction of the Radial Direction}
\label{subsec:mi-mechanism}

In all of the accreting cases the stress that reaches threshold is not the
tidal Weyl field of Eq.~\eqref{eq:schwT}. It is the trace-free Ricci stress of
the counter-streaming fluxes themselves, and the difference is enormous.

Throughout this section we write the Kerr radial coordinate in the
inner-horizon region as
\begin{equation}
	r \;=\; r_{-} + r_* ,
	\label{eq:rstar}
\end{equation}
with $r_*$ the displacement above the inner horizon. All of the relevant
structure lies at $r_* \ll r_{-}$: the inflating layer occupies
$r_* \lesssim d$, and $\BQ$ forms at the particular value of $r_*$ given
below. Setting $r \to r_{-}$ in coefficients is accurate to $r_*/r_{-}$,
which is at most a part in $10^{6}$ in every case considered.

Consider a locally orthonormal frame at $r\approx r_{-}$ in which ingoing and
outgoing null streams cross. For balanced streams the effective stress tensor
is the sum of two null dusts propagating along $\pm\hat r$,
\begin{equation}
	T_{\hat\mu\hat\nu} \;\simeq\; u\operatorname{diag}(1,1,0,0)
	\quad\text{in }(t,r,\theta,\phi),
	\label{eq:nulldust}
\end{equation}
where $u$ is the center-of-mass energy density. This tensor is trace free,
so $R=0$ and $S_{\hat\mu\hat\nu}=R_{\hat\mu\hat\nu}=(8\pi G/c^4)\,u
\operatorname{diag}(1,1,0,0)$. Writing
\begin{equation}
	\mathcal{T}_{\rm MI} \;\equiv\; \frac{8\pi G\,u\,q^2}{c^4} ,
	\label{eq:TMI}
\end{equation}
Eq.~\eqref{eq:ricciresponse} gives the deformed element, in the spatial triad of the center-of-momentum frame with $\hat1$ along the streaming axis,
\begin{equation}
	H_{ij} =
	\operatorname{diag}\!\left(1-C_R\mathcal{T}_{\rm MI},\,1,\,1\right)_{(\hat1,\hat2,\hat3)}
	+ \chi^Q_{ij}(q,t).
	\label{eq:miH}
\end{equation}
The triad labels are frame directions, not Boyer--Lindquist coordinates: the
streaming axis $\hat1$ is spacelike in the center-of-momentum frame by
construction, since the two momenta cancel there, even though it points along
increasing $r$ and $\partial_r$ is timelike between the horizons. The timelike
leg of the frame is the four-velocity of the observer for whom the streams
balance.

Equation~\eqref{eq:miH} is the physical heart of the mechanism, and it is
worth stating in words. The counter-streaming is directed in opposite senses
along a single axis, and the stress it produces is concentrated on that axis
alone. The deformation does not squeeze the element uniformly; it drives one
eigenvalue, the one along the streaming axis, toward zero while leaving the
transverse directions untouched. Signature failure here is the extinction of the radial
direction, precisely the direction along which the two streams run through
each other. Once $1-C_R\mathcal{T}_{\rm MI}+\chi^Q_{\hat1\hat1}$ crosses zero the element has two
timelike directions, the light cone has opened into a connected set, and the
element is gone.

The threshold in $C_R\mathcal{T}$ for this pattern has been scanned directly with the
same gauge-reduced six-metric ensemble used for the isotropic case, applying
the Ricci response $s^{(R)}_{\hat\mu\hat\nu}=-C_Rq^2S_{\hat\mu\hat\nu}$ to the
counter-streaming pattern of Eq.~\eqref{eq:nulldust} in the local
center-of-momentum frame and locating the loss of a periodically winding
network of complete $3q$ Lorentzian patches. The result is
$\Tqb = 1.00\pm0.03$ over $\alpha = 1.9$ to $10$; in density,
\begin{equation}
	\begin{split}
	\frac{\rho_{\rm qb}}{\rho_P} &\;\simeq\; \frac{1}{8\pi\alpha^2 C_R}\\
	&\;=\; 4.2\times10^{-2}
	\left(\frac{\alpha}{2.5}\right)^{-2} ,
	\end{split}
	\label{eq:rho_qb_MI}
\end{equation}
or $4.3\times10^{-3}$ at $\alpha=10$ and $7.3\times10^{-2}$ at $\alpha=1.9$. Here and in every numerical value quoted below we assume $C_R=0.15$; the symbolic forms retain $C_R$, so a different choice rescales each result as shown. The scans are run with the response coefficient set to unity, so $\Tqb$ is a threshold in the metric deformation itself and the adopted $C_R$ enters only in converting that deformation back to a density or an interior mass. The same holds for the radii tabulated below. Both coefficients inherit the truncation of the normal-coordinate expansion at second order, applied at precisely the displacement where the deformation reaches order unity; they should be read as fixing the scale of the conversion rather than its third digit, and every absolute scale quoted here accordingly carries its coefficient explicitly.

That $\Tqb$ sits so close to unity, and so nearly independently of $\alpha$,
is a consequence of the eigenvalue pattern rather than of the fluctuation
statistics. Because the stress is concentrated on the radial axis alone, the
deterministic part of the local metric is
$\operatorname{diag}(-(1+C_R\mathcal{T}),1,1,1-C_R\mathcal{T})$ in this frame, and
its radial eigenvalue passes through zero at $C_R\mathcal{T}=1$ with the quantum field switched
off entirely: the metric is degenerate there, $\det g = 0$, with a
one-dimensional null direction along the streaming axis. The vacuum
fluctuation neither creates nor prevents that crossing; it only displaces the
density at which a connected winding network is finally lost, by $+3\%$ at
$\alpha=2.5$ and $\alpha=10$ and by $-1\%$ at $\alpha=1.9$, where the larger
amplitude $\Delta_g=1.596/\alpha$ carries elements across ahead of the
background. This is the opposite limit from the isotropic radiation case,
where the pattern depresses each spatial eigenvalue by only $C_R\mathcal{T}$ and
the fluctuations do most of the work, $\Tqb=0.685$ at $\alpha=2.5$
\citep{shaya2026signature}. The mass-inflation
configuration is thus the one case treated here in which $\BQ$ is located to
a few percent by the classical stress alone, and $\rho_{\rm qb}$ is fixed by
the patch size and the response coefficient.

The mass-inflation layer is built from crossed null fluxes, and the blueshift
that drives $m(v)$ acts on null geodesics, so the stress above is traceless by
construction.

It remains to express $u$ in terms of the amplification of the interior mass.

\emph{The density.} The Misner--Sharp mass function ties the radial gradient
of the interior mass to the local energy density, $\partial m/\partial r =
4\pi r^2 u/c^2$. Writing $d \equiv m/|\partial m/\partial r|$ for the scale on
which the mass function varies, this is the identity
\begin{equation}
	u \;\simeq\; \frac{m c^2}{4\pi r_{-}^2\,d} .
	\label{eq:sheet}
\end{equation}
No proper thickness enters: $d$ is the logarithmic variation scale of $m$ in
areal radius, which is the quantity the growth law supplies.

\emph{The growth law.} Hamilton and Avelino \citep{hamiltonavelino2010} give
the growth of the counter-streaming boost $\beta$, the relative Lorentz factor
of the ingoing and outgoing streams, as $d\ln\beta/d\ln r = -\mathcal{N}/\mu$,
with $\mu$ the dimensionless accretion rate, the light-crossing time of the
hole divided by its accretion time, and $\mathcal{N}\equiv\lambda^2/2C^2 \approx 28$ fixed by the numerical constants
$\lambda$ and $C$ of their inflationary solution. During strong inflation $m \simeq r\beta^2/2$
in geometrized units, so $d\ln m/d\ln r = 1 + 2\,d\ln\beta/d\ln r$ and the
amplification per $e$-fold of radius is $|d\ln m/d\ln r| \simeq
2\mathcal{N}/\mu$, whence
\begin{equation}
	d \;=\; \frac{\mu\,r_{-}}{2\mathcal{N}} ,
\end{equation}
derived rather than imposed. In advanced time the same growth reads $m(v)
\propto e^{\kappa_{-}v}$ with $\kappa_{-}$ the inner-horizon surface gravity;
the two statements differ only in whether the growth is measured against $v$
or against $\ln r$, and $\mu$ is what relates them.

\emph{In proper time.} Along an infalling worldline the exponential appears as
$\rho \propto \tau^{-2}$, with $\tau$ the proper time remaining, so
$\tau \propto e^{-\kappa_{-}v/2}$ and the boundary is reached with
\begin{equation}
	\tau_{\rm qb} \;\simeq\; \frac{r_{-}}{c}\,\nBQ^{-1/2}
	\label{eq:tauqb}
\end{equation}
still to run. For the cases of Table~\ref{tab:mu-range} this is $10^{5}$ to
$10^{20}$ Planck times, against the $3q/c$ that a patch requires. The element
that fails at $\BQ$ is therefore a four-dimensional region with proper time to
spare, not a degenerate slice.

The classical solution places no floor under $d$. The framework does: no
structure exists on a scale finer than one spatial quantum, so the stress is
evaluated across $q$ and
\begin{equation}
	d \;=\; \max\!\left(\frac{\mu\,r_{-}}{2\mathcal{N}},\; q\right),
	\qquad
	\mu_q \;\equiv\; \frac{2\mathcal{N} q}{r_{-}} .
	\label{eq:dlayer}
\end{equation}
For a $10\,M_\odot$ hole at $a/r_g\sim0.7$, $\mu_q \simeq 6\times10^{-37}$.
Above that rate the layer is thicker than a quantum and the growth law sets
$d$; below it, which includes the quiescent late life of an
isolated hole, the classical solution demands a sheet thinner than $\lP$ and
the quantum floor takes over. The two regimes are branches of a single
expression.

Combining Eqs.~\eqref{eq:TMI}--\eqref{eq:dlayer} with $m = nM$ and
$Gm/c^2 = n r_g$,
\begin{equation}
	\mathcal{T}_{\rm MI} \;=\; \frac{2\,n\,r_g\,q^2}{r_{-}^2\,d} ,
	\label{eq:TMIn}
\end{equation}
to be compared with the tidal strength of the same interior mass at the same
radius, $\mathcal{T}^{(W)}_{\rm Kerr} = n r_g q^2/r_{-}^3$, the same pattern
as Eq.~\eqref{eq:schwT} evaluated at $r_{-}$ with the inflated mass.
Their ratio is
\begin{equation}
	\frac{\mathcal{T}_{\rm MI}}{\mathcal{T}^{(W)}_{\rm Kerr}} \;=\; \frac{2r_{-}}{d} ,
	\label{eq:ratio}
\end{equation}
which runs from $\sim10^{8}$ at peak collapsar fallback to $\sim10^{38}$ at
the quantum floor. The two stresses respond to different things. The Weyl tidal field at $r_{-}$ depends only on
the mass enclosed, and so is insensitive to how that mass is distributed,
while the trace-free Ricci stress responds to the local energy density, which
mass inflation drives up by confining the amplified mass to a layer of
thickness $d$. The streams' own pressure therefore reaches the threshold long
before the tidal field of the inflated mass becomes relevant, and the ratio of
the two is the factor $2r_{-}/d$ of Eq.~\eqref{eq:ratio}.

Setting $C_R\mathcal{T}_{\rm MI} = \Tqb$ caps the internal mass parameter
$n\equiv m_{\max}/M$ at
\begin{equation}
	\nBQ \;=\; \frac{\Tqb\,r_{-}^2\,d}{2\,C_R\,r_g\,q^2},
	\label{eq:nmax}
\end{equation}
or, in the two regimes and in the form that separates the spin and mass
dependence,
\begin{equation}
	\nBQ =
	\begin{cases}
	\dfrac{\Tqb\,\mu}{2\mathcal{N}C_R\alpha^2}
	\left(\dfrac{r_{-}}{r_g}\right)^{3}\left(\dfrac{r_g}{\lP}\right)^{2},
	& \mu > \mu_q,\\[3.5ex]
	\dfrac{\Tqb}{2C_R\alpha}
	\left(\dfrac{r_{-}}{r_g}\right)^{2}\left(\dfrac{r_g}{\lP}\right),
	& \mu < \mu_q .
	\end{cases}
	\label{eq:nmaxcases}
\end{equation}
The two branches agree at $\mu=\mu_q$. The upper branch scales as
$(r_{-}/r_g)^3(r_g/\lP)^2$ times the factor
$\Tqb\mu/2\mathcal{N}C_R\alpha^2$, which carries the accretion rate: the
boundary is not a property of the hole alone.

\subsection{Case 1: The Isolated Kerr Black Hole, No Accretion}
\label{subsec:case-vacuum}

The first case is a black hole that forms by collapse and thereafter accretes
nothing. In the original treatment of mass inflation by
Poisson and Israel \citep{poissonisrael1990}, collapse itself generates a
decaying Price tail of outgoing radiation, part of which is backscattered by
the curvature of the hole, so ingoing and outgoing streams are present at the
inner horizon even in perfect isolation. Their relativistic counter-streaming
ignites the mass-inflation instability, in which the streaming pressure and
energy flux source a gravitational force that accelerates the streams further
through each other, in opposite senses for ingoing and outgoing frames; the
exact mass-inflation solution of Ori \citep{ori1991} is constructed for this
case. The divergence is obtained by solving the field equations with both
streams present as sources, not by superposing the solutions for each stream
separately, so the interaction between them is part of the derived result
rather than an artifact of an additive construction. Nor does even the Price tail exhaust the supply: the renormalized Unruh
fluxes $\langle T_{uu}\rangle$ and $\langle T_{vv}\rangle$ at the Kerr Cauchy
horizon are generically nonvanishing and comparable in magnitude
\citep{zilberman2022kerr,mcmaken2024}, so a crossflow of the pattern of
Eq.~\eqref{eq:nulldust} is present in any Kerr geometry whatever its history. Hamilton and Avelino \citep{hamiltonavelino2010} show that the growth
rate scales inversely with the dimensionless accretion rate $\mu$, the
light-crossing time divided by the accretion time, so that the instability is
most violent when the source is most feeble. The interior mass, the center-of-mass energy density, and the
Weyl curvature all exponentiate together while the radius barely moves.

This is the case in which the layer becomes Planck thin. Radiative tails from
spherical collapse fall off as inverse power laws in advanced time
\citep{poissonisrael1990}, so by the time the interior mass has amplified
appreciably the tail has decayed far below any realistic accretion $\mu$, and
in particular below $\mu_q$. The lower branch of Eq.~\eqref{eq:nmaxcases}
applies: $d$ sits on its floor, the entire inflated mass of the interior is
concentrated in a shell one spatial quantum thick, $q = 2.5\,\lP$, and it is
that concentration that ends the manifold. For a $10\,M_\odot$ hole at
$a/r_g\sim0.7$ this gives $\nBQ \sim 9\times10^{37}$, or $\ln\nBQ \approx 88$.

Two quantities characterize the formation of $\BQ$ and apply unchanged
throughout the rest of this section. The first is $\nBQ$ of
Eq.~\eqref{eq:nmax}. The second is the radius at which $\BQ$ forms. The
available amplification per $e$-fold of radius is $2\mathcal{N}/\mu$, so
obtaining $\ln \nBQ$ folds costs
\begin{equation}
	\rBQ - r_{-} \;\approx\;
	\frac{\mu\,r_{-}}{2\mathcal{N}}\,\ln \nBQ
	\;=\; 0.018\,\mu\,r_{-}\ln\nBQ ,
	\label{eq:dlnr}
\end{equation}
with the fitted $\lambda \approx 3$, $C \approx 0.4$ of
\citep{hamiltonavelino2010}. These two parameters are measured from
self-similar models of a \emph{charged} spherical hole with $Q_\bullet/M_\bullet
= 0.8$, and we carry them over to the rotating case on the standard surrogacy
of charge for angular momentum
\citep{poissonisrael1990,hamiltonavelino2010}; they enter
Eq.~\eqref{eq:dlnr} only through the combination
$2C^2/\lambda^2 \approx 0.036$, so the conclusion is insensitive to their precise values.

For the isolated hole, $\mu$ is set by the Price tail rather than by external
accretion, so the boundary forms at $r_{-}$ to extraordinary precision, and in
fact more tightly frozen than in any case in which external material feeds the
interior. What the classical literature describes as a weak null singularity
on the Cauchy horizon is preempted: the manifold terminates at $\BQ$ while
every curvature invariant is still finite, and the null singularity is not part
of it.

The spherical scale of Eq.~\eqref{eq:rBQnum} accordingly serves throughout the
paper as a reference length rather than as the location of the boundary in any
physical case.

\subsection{Case 2: Realistic Accretion Onto a Rapidly Spinning Black Hole}
\label{subsec:case-kerr-inflation}
\label{subsec:kerr-mass-inflation}

Astrophysical black holes carry angular momentum, and unless the spin is
unusually small, the inner Cauchy horizon $r_{-}$ sits at a macroscopic radius,
some fraction of $r_g$. This is the realistic case, covering the black hole's
entire life from a few crossing times after formation onward, including the
vigorous fallback epoch.

What distinguishes this case from Case 1 is only the value of $\mu$, the
streams now being supplied by accretion rather than by the Price tail. Since
every realistic $\mu$ exceeds $\mu_q$, the upper branch of
Eq.~\eqref{eq:nmaxcases} applies and $\nBQ$ is proportional to $\mu$.

\begin{table}[ht]
\centering
\small
\begin{tabular*}{\linewidth}{@{\extracolsep{\fill}}|lllll@{\hspace{2mm}}|}
	\hline
	epoch & supply & $\mu$ & $\ln\nBQ$ & $\rBQ-r_{-}$ (m) \\
	\hline
	peak collapsar & $0.1\,M_\odot$/s & $5(-7)$ & 157 & $8(-3)$\\
	one day & $1\,M_\odot$/d & $6(-11)$ & 148 & $4(-7)$ \\
	tail & $0.1\,M_\odot$/3mo & $6(-14)$ & 141 & $4(-10)$ \\
	quiescent & $M/t_H$ & $1(-22)$ & 121 & $8(-19)$ \\
	CMB floor & $\pi r_g^2u_\gamma/c$ & $2(-49)$ & 88 & $<q$ \\
	\hline
\end{tabular*}
	\caption{Amplification cap and displacement $\rBQ-r_{-}$ of $\BQ$ above the inner horizon across the
	accretion history of a $10\,M_\odot$ hole at $a/r_g = 0.7$, for which
	$r_{-}/r_g = 0.286$. The dimensionless rate is
	$\mu = G\dot M/c^3$, or $4.93\times10^{-6}$ per
	$M_\odot\,{\rm s}^{-1}$, independent of hole mass when expressed
	through $\dot M$. Values use Eq.~\eqref{eq:nmaxcases} with
	$\Tqb=1.00$, $C_R=0.15$, $\alpha=2.5$, and Eq.~\eqref{eq:dlnr}. The bottom row
	lies below $\mu_q\simeq6\times10^{-37}$ and therefore uses the
	quantum-floor branch, for which $\rBQ-r_{-}$ falls below one spatial quantum, $q=4\times10^{-35}$~m, and is not a resolvable displacement. Entries $a(b)$ denote $a\times10^{b}$.}
\label{tab:mu-range}
\end{table}

\noindent
For a $10\,M_\odot$ hole, whose light-crossing time is $GM/c^3 \simeq
5\times10^{-5}$~s: peak collapsar fallback at $\sim
0.1\,M_\odot\,\mathrm{s}^{-1}$ corresponds to $\mu \sim 5\times10^{-7}$,
fallback of order $1\,M_\odot$ over a day to $\mu \sim 6\times10^{-11}$,
a tail delivering $\sim 0.1\,M_\odot$ over months to $\mu \sim
6\times10^{-14}$, and accretion of its own mass over a Hubble time to
$\mu \sim 10^{-22}$. Even that is not a floor: an isolated hole still sits in
the cosmic microwave background, whose Bondi capture supplies $\dot M \sim
\pi r_g^2 u_\gamma c^{-1}$ and gives $\mu \sim 2\times10^{-49}$ for a
$10\,M_\odot$ hole today, falling as $(1+z)^4$ with cosmic expansion. Across
the entire range, from a merger transient down to the CMB floor, $\BQ$ forms
at $r_{-}$ to within parts per million or better, and, throughout most of the
hole's life, to within parts per $10^{22}$. The freezing of the radius is a
property of the whole regime $\mu \ll 1$ and not of any particular epoch. In a
realistic rotating black hole $\BQ$ is therefore a macroscopic surface, and the
microscopic radius of Eq.~\eqref{eq:rBQnum} is never approached.

The $\mu$-dependence of $\nBQ$ is a genuine prediction of the metric-stress
criterion. Slower accretion produces a thinner and more violently inflating
layer, which reaches the stress threshold at a smaller total amplification.
The interior of a quiescent hole terminates with less inflated mass than that
of a vigorously fed one, even though both terminate at the same radius. The
same trend appears in the semiclassical calculation of McMaken and Hamilton
\citep{mcmakenhamilton2022}, who find the renormalized vacuum polarization
$\langle\phi^2\rangle_{\rm ren}$ over the inflationary Kasner geometry to be
larger for smaller accretion rates and smaller spins.

A boundary reached at finite amplification also removes the standard objection
that mass inflation depends on the infinite future. The classical Cauchy
horizon is generated by the $v\to\infty$ rays, and the divergence of $m(v)$ is
attained only in that limit, so an evaporating hole, whose exterior ends at
finite $v$, never supplies them. The quantum boundary requires no such limit.
With $\epsilon \equiv \sqrt{1-a^2/r_g^2}$ and
$\kappa_{-} = \epsilon/[2r_g(1-\epsilon)]$, reaching $\ln\nBQ$ takes an
advanced time
\begin{equation}
	v_{\rm qb} \;=\; \frac{\ln\nBQ}{\kappa_{-}}
	\;=\; \frac{2r_g(1-\epsilon)\ln\nBQ}{\epsilon\,c} ,
	\label{eq:vqb}
\end{equation}
which for a $10\,M_\odot$ hole at the Thorne limit $\epsilon \simeq 0.063$ is
$0.13$~s at $\ln\nBQ = 88$ and $0.23$~s at $157$. The exterior can supply
$\kappa_{-}t_{\rm evap} \sim 2\times10^{80}$ e-foldings before evaporating;
the boundary needs of order $10^{2}$. Equating $v_{\rm qb}$ to $t_{\rm evap}$
would require $\epsilon \sim 10^{-80}$, far below the Thorne limit and below
any spin a hole could hold. Even in the quiescent case, with no infall beyond
the cosmic microwave background, the influx needed to form $\BQ$ arrives
within a second of horizon formation, and everything afterward in the hole's
history happens to a spacetime that already terminates at $r_{-}$.

The classical divergence is unbounded, and every treatment since Poisson and
Israel has looked for the effect that terminates it. Carballo-Rubio \emph{et
al.} \citep{carballorubio2024} find a classical regulator in the feedback of a
dynamically evolving inner trapping horizon, but their numerical solution
tracks the stationary prediction through some $60$ decades of growth, $\ln M
\approx 138$, before it acts; $\ln\nBQ$ lies at or below that for most of a
hole's life, so the regulator has not begun to operate where the stress
threshold is met and Eq.~\eqref{eq:nmax} is unaffected. The semiclassical
route \citep{zilberman2022kerr,mcmaken2024,mcmakenhamilton2022} establishes
that the renormalized stress diverges at the Cauchy horizon of a spinning hole
even without accretion, but it is explicitly valid only while the geometry
remains approximately vacuum Kerr, and it hands the final evolution to a theory
of quantum gravity it does not supply. What all of these analyses require, and
none provides, is a criterion that says where the geometry ceases to exist.
$\BQ$ is that criterion, and it is met while the classical description is still
the accurate one.

Carballo-Rubio \emph{et al.} \citep{carballorubio2024} also adopt a closely
related endpoint criterion, proposing that the physically meaningful
definition of mass inflation is a transient exponential buildup persisting
until Planckian conditions are reached, rather than a mathematical divergence.
They stop at that point, observing that generic black holes with non-extremal
inner horizons ``cannot be the endpoint of a stellar collapse'' and that
determining that endpoint remains an open question. The $\BQ$ boundary supplies
one: the manifold terminates where the metric-stress criterion is met, and the
question of what geometry lies beyond does not arise.

\subsection{Case 3: Slow Infall With Low $J/M$ --- The Near-Spherical Interior}
\label{subsec:case-schwarzschild-collapse}
\label{subsec:schwarzschild-bq}

For a weakly spinning hole, the inner horizon lies at $r_{-} \approx a^2/2r_g$,
small compared with $r_g$ but, as noted above,
still enormous compared with $\rBQ$ for any spin above $a/r_g \sim 2\times10^{-13}$
for a $10\,M_\odot$ hole. Infalling matter is therefore not delivered to the
center; it encounters an inner horizon first.

What is less obvious is that the counter-streaming instability operates even
when the spin is barely sufficient to produce an outgoing stream. Weak
centrifugal repulsion makes the outgoing component very much smaller than the
ingoing one, but Hamilton and Avelino \citep{hamiltonavelino2010} show that
very unequal streams still inflate: the weaker stream undergoes a
pre-inflationary acceleration until the product of its accretion rate and
velocity matches that of the dominant stream, whereupon inflation proceeds with
a growth rate set by the larger rate. A small outgoing admixture suffices.

Equations~\eqref{eq:nmaxcases} and~\eqref{eq:dlnr} apply here with $r_{-}(a)$
substituted for its high-spin value. On the upper branch $\nBQ \propto
(r_{-}/r_g)^3$, so the cap falls as $a^6$; on the quantum-floor branch it
falls only as $a^4$. The crossover rate $\mu_q = 2\mathcal{N}q/r_{-}$ rises as
the inner horizon shrinks, so weakly spinning holes enter the floor regime at
higher accretion rates: for $a/r_g \sim 10^{-6}$ and a $10\,M_\odot$ hole,
$r_{-}\simeq7\times10^{-9}$~m and $\mu_q \simeq 3\times10^{-25}$, so quiescent
accretion at $\mu\sim10^{-22}$ still lies on the upper branch and gives
$\nBQ \sim 10^{16}$. The radius contraction is unaffected: by
Eq.~\eqref{eq:dlnr}, $\rBQ-r_{-}$ is proportional to $\mu\ln\nBQ$ and is
tiny at any spin. Thus $\BQ$ forms at $r_{-}(a)$, at a radius that
interpolates continuously between the two limits: as $a \to 0$, $r_{-}$ descends
toward $\rBQ$ and $\BQ$ approaches the spherical reference surface of
Eq.~\eqref{eq:rBQnum}; as the spin grows, $\BQ$ moves outward toward the
macroscopic radius of Case 2.

\subsection{Case 4: Mergers and the Assembly Event}
\label{subsec:case-merger}

The remaining case is that of accretion so rapid that the ordering of events
reverses. Mass inflation is self-limiting: as the interior mass grows it
eventually contributes more to the gravitational force than the streaming
pressure does, at which point the counter-streaming velocity peaks and the
streams collapse toward zero radius \citep{hamiltonavelino2010}. Reaching that
peak costs a radial contraction $|\Delta\ln r| \approx 1/2$. Comparing this with
the contraction needed to reach the stress threshold, Eq.~\eqref{eq:dlnr},
defines a critical accretion rate $\mu_{\rm crit}$ above which the collapse
phase preempts the boundary:
\begin{equation}
	\mu \;\gtrsim\; \mu_{\rm crit} \;\sim\; \frac{\mathcal{N}}{\ln \nBQ}
	\;\sim\; 2\times10^{-1}.
	\label{eq:mucrit}
\end{equation}
Below $\mu_{\rm crit}$ the stress threshold arrives while the radius is still
essentially $r_{-}$ and $\BQ$ halts the inflation there; above it, the velocity
peaks first and the streams run inward while still below threshold. The
self-similar models of \citep{hamiltonavelino2010} straddle this value: their
fitted stream parameters are $\mu = 0.074$, $0.238$ and $0.767$, so the first
is halted near $r_{-}$ while the last two contract to a small radius before the
threshold is met. Those models sit near $\mu_{\rm crit}$ because that is the
range that is numerically tractable, not because it is the range a black hole occupies; no
epoch in the life of an astrophysical hole reaches it by accretion, since even
peak fallback falls some six orders of magnitude short.

Attaining $\mu \sim \mu_{\rm crit}$ requires assembling of order the hole's own
mass within $(GM/c^3)/\mu_{\rm crit} \sim 10^{-3}$~s, which is the dynamical
formation of the hole itself or a merger, not accretion of any kind. For
material caught in such an event, the streams do plunge. Once they do, the
thin-layer geometry that produced Eq.~\eqref{eq:TMIn} is gone: the flow is no
longer a shell at a stationary radius but a converging plunge, and the
tidal deformation of Eq.~\eqref{eq:schwT} evaluated with the inflated interior
mass again becomes the operative stress. The boundary therefore forms deep in
the interior, near the spherical reference radius but computed with $m = nM$
rather than $M$, so that by Eq.~\eqref{eq:rBQ} it sits a factor $n^{1/3}$
farther out than the naive estimate.

The amplification attained before the velocity peak follows from
Eq.~\eqref{eq:dlnr} with $|\Delta\ln r| = 1/2$: $\ln n \approx
2\mathcal{N}/\mu \approx 56/\mu$, which for a merger transient in the range
$\mu \sim 0.3$--$1$ gives $n$ from $\sim 10^{76}$ down to $\sim 10^{24}$. For a
$10\,M_\odot$ hole, whose vacuum boundary would lie at $2.2\times10^{-22}$~m,
this places $\BQ$ between roughly $3\times10^{-14}$~m, still microscopic but
eight orders of magnitude above the value obtained by ignoring inflation
altogether, and $r_{-}$ itself. The range is wide because $n$ depends
exponentially on $\mu$, and it is bounded above by the marginal case: the
boundary reaches $r_{-}$ already at $\mu \simeq 0.3$, and as $\mu \to
\mu_{\rm crit}$ from above it returns continuously to $r_{-}$, joining Case 2. The chaotic Kasner dynamics of such a plunge are treated in
Sec.~\ref{subsec:bkl-suppression}.

This case is a transient, not a final state. A merger remnant rings down quickly: the
fundamental quadrupole quasinormal mode has damping time $\tau = GM/c^3y$ with $y\approx0.09$, about $11$ crossing times, and the
signal is described by perturbation theory from some $10\,GM/c^3$ after
merger \citep{ligo2017basic,bhagwat2020}. Within a few hundred crossing times,
tens of damping times, the remnant has settled to a quiet Kerr hole of roughly twice the
progenitor mass and high spin. It therefore possesses a macroscopic inner
horizon of its own, and everything falling in thereafter meets the boundary of
Case 2. The deep-interior $\BQ$ of Case 4 is reached only by material present
during the assembly event itself, over the few hundred crossing times before
the remnant settles; the surface encountered by ordinary infall over the whole
of the hole's subsequent life is that of Case 2.

\subsection{The GHY Action at the $\BQ$ Boundary}
\label{subsec:ghy-action}

The Einstein--Hilbert action requires the Gibbons--Hawking--York (GHY)
term \citep{gibbonshawking1977},
\begin{equation}
	S_\mathrm{GHY} = \frac{c^3}{8\pi G}\int_{\partial\mathcal{M}} \!d^3x\,
	\epsilon\,\sqrt{|h|}\,K,
	\label{eq:ghy-def}
\end{equation}
where $K$ is the trace of the extrinsic curvature of the boundary and
$\epsilon = -1$ on spacelike segments. When $\BQ$ introduces an inner
boundary, the action acquires a second GHY contribution from that surface.
This is the same bookkeeping that, at the microscopic level, attaches a
boundary term to every individual excision: $\BQ$ is the macroscopic limit in
which those excisions have merged into one closed surface.

In all four cases $\BQ$ lies inside a horizon, where $r$ is timelike, and is
therefore a spacelike terminal hypersurface (Sec.~\ref{subsec:signature}). In
every case, then, $\epsilon = -1$ and the integrand $\epsilon\sqrt{h}\,K$
evaluated on a slice of constant $r$ has the finite form obtained when
the divergent factors in $\sqrt{h}$ and $K$ cancel:
\begin{equation}
	\epsilon\,\sqrt{h}\,K \;=\; \tfrac{1}{2}\,c\,(3r_s - 4\rBQ)\sin\theta,
	\label{eq:ghy-integrand}
\end{equation}
with $r_s = 2Gm/c^2$ the Schwarzschild radius of the quasi-local mass $m$ at
the boundary. Integrating over a spatial slice of duration
$\Delta t$ gives the boundary action per slice,
\begin{equation}
	\begin{split}
	S_\mathrm{GHY}^{\rm qb}
	&\;=\; \frac{c^3}{4G}\,(3r_s - 4\rBQ)\,\Delta t\\
	&\;\approx\; \frac{3}{2}\,m c^2\,\Delta t,
	\end{split}
	\label{eq:ghy-slice}
\end{equation}
using $\rBQ \ll r_s$ in the second form, where $m$ is the mass characterizing
the geometry local to the boundary, identified below. Spin enters through an
order-unity coefficient that vanishes in the extremal limit
($\kappa_{-} \to 0$); for realistic $a/r_g \sim 0.7$ the coefficient is of
order the $3/2$ shown, and we do not track it further.

The mass appearing here is the quasi-local mass at the boundary, the
Misner--Sharp mass in the spherical models used as surrogates for the rotating
case, and in Cases 1--3 that is the inflated $m = \nBQ M$ of
Eq.~\eqref{eq:nmax}, not the exterior $M$. The two are both physical and
describe different regions. This quasi-local mass is what sets the local
metric stress: it is precisely the mass whose confinement to a layer of
thickness $d$ makes $\mathcal{T}_{\rm MI}$ of Eq.~\eqref{eq:TMIn} reach
$\Tqb$ and thereby fixes the location of $\BQ$, while the ADM mass $M$
characterizes the exterior geometry and is unaffected because the inflating
region lies inside the horizon and cannot signal to infinity. The energy
difference resides in the relative kinetic energy of the counter-streaming
flows, whose center-of-mass density exponentiates even though each stream's
proper density does not. Since the GHY term is a local integral over the
boundary, built from the extrinsic curvature of that surface, it is the local
mass that enters, and consistency requires it: the same $m = \nBQ M$ that
places $\BQ$ through Eq.~\eqref{eq:nmax} must govern the action of the surface
it places. Thus
\begin{equation}
	S_\mathrm{GHY}^{\rm qb} \;\approx\; \tfrac{3}{2}\,\nBQ \,M c^2\,\Delta t .
	\label{eq:ghy-slice-n}
\end{equation}

The full covariant boundary component of $\partial\mathcal{M}$ is the
three-surface swept out by $\BQ$ over the hole's existence
$\tau_\mathrm{evap}$. Integrating Eq.~\eqref{eq:ghy-slice} at fixed $M$
gives
\begin{equation}
	S_\mathrm{GHY}^{\rm qb,\,\mathrm{life}}
	\;\approx\; \tfrac{3}{2}\,\nBQ\,M c^2\,\tau_\mathrm{evap},
	\label{eq:ghy-life}
\end{equation}
or $\sim 10^{156}\,\hbar\,\nBQ\,(M/M_\odot)^{4}$, using
$\tau_\mathrm{evap} \sim (M/M_\odot)^3 \times 10^{67}\,\mathrm{yr}$
for Hawking evaporation. With $\nBQ$ in the range of
Table~\ref{tab:mu-range}, $|S_\mathrm{GHY}^{\rm qb}| \gg \hbar$ by an
overwhelming margin, so the terminal surface lies deep in the semiclassical
regime and a stationary-phase evaluation of the integral over geometries is
justified there. This is not by itself a stability result: whether the boundary
is stable against a given class of perturbations is fixed by the second
variation of the full action under those perturbations, not by the magnitude of
$S_\mathrm{GHY}^{\rm qb}$, and we do not attempt that calculation here. The
argument that the boundary holds its position, given in
Sec.~\ref{subsec:schwarzschild-radius}, is a statement about the support of
$\Psi$ on either side of $\rBQ$ and is independent of the action's magnitude.
The rate, of order $\nBQ Mc^2$ per unit time, is set by the quasi-local mass at
the boundary. Nothing lies inside to be enclosed, and none is needed: the
Misner--Sharp mass is determined by the areal radius and by
$\nabla_\mu r\,\nabla^\mu r$ on the surface itself, so it remains well defined
with no interior. The boundary term performs the bookkeeping that the excised
region would otherwise have done, registering that mass on the terminal
surface.

An immediate question is whether this internal boundary should contribute
to the Hawking spectrum. The standard derivation quantizes fields on a
one-boundary geometry, the outer horizon, and traces over the modes
that fall behind it, producing thermal emission at $T_H = \hbar \kappa_{+} /
2\pi c k_B$ with $\kappa_{+}$ the surface gravity of the outer horizon.
With $\BQ$ present, the interior is bounded, and the modes traced over are
those that fall between the two boundaries rather than those that fall
``into infinity.'' Two features of this modification deserve emphasis.

First, the outer emission spectrum is unchanged. The Bogoliubov coefficients
that produce the thermal flux at $\mathcal{I}^+$ depend on the mode
propagation between the collapsing surface and $\mathcal{I}^{+}$, and this
propagation is unaffected by whatever terminal condition is imposed inside
$r_{+}$. The presence of $\BQ$ prunes the states of the interior partner
modes but not the emission itself; $T_H$ is set by $\kappa_{+}$ and remains
$T_H = \hbar c^3 / 8\pi G M k_B$ up to the standard spin correction, with the
exterior ADM mass $M$ appearing here rather than the local $\nBQ M$ of
Eq.~\eqref{eq:ghy-slice-n}, since the outer horizon lies in the exterior
geometry.

Second, the $\BQ$ surface does not radiate. It is spacelike, not a horizon:
it has no null generators, so no surface gravity is defined on it, and the
construction that assigns a temperature to a Killing horizon
\citep{gibbonshawking1977} has nothing to act on. Nor is there a region beyond
it whose modes could be traced out, or a future into which a flux could
propagate. The GHY term registers the existence of the terminal boundary in
the action, but no thermal emission accompanies it. Whatever thermal character
the interior has is that of the outer horizon alone.

What does change is the entropy bookkeeping. The interior partner modes of
the outer flux are cut off at $\BQ$ rather than extending to a Cauchy horizon
or a singularity, so the region available to them has finite proper volume;
and since no element smaller than $q_{\min}$ exists, that region contains a
finite number of independent metric elements. The number of degrees of freedom
available to the partners is therefore finite, and the entanglement entropy of
the Hawking radiation is bounded above by that count rather than growing
without limit. That
is a separate calculation, but it is the natural place where the
information-loss question would be reopened in this framework.

\subsection{BKL Suppression and Symmetry Annihilation} \label{subsec:bkl-suppression}

The collapsing interior undergoes chaotic, anisotropic Mixmaster epochs \citep{bkl1970}, and oscillating Kasner transitions with increasing frequency as $r \to 0$. In rotating holes this collapse phase follows mass inflation \citep{hamilton2017}, so the mechanism below applies to the plunging Case 4 interiors, to non-rotating holes, and to matter-filled interiors that reach the collapse phase.

Semiclassical backreaction is argued to carry the local geometry through mass
inflation and then, once the radial collapse stalls, into inflating Kasner
bounces
toward a chaotic spacelike singularity \citep{mcmaken2024}. Those bounces
require the radial collapse to become of order unity, $|\Delta\ln r|\sim1$,
hence $\ln n \sim 2\mathcal{N}/\mu$: some $10^{8}$ e-foldings for the
peak-collapsar case and $10^{50}$ at the CMB floor. The boundary is reached at
$\ln\nBQ = 88$--$157$, five to fifty orders of magnitude in amplification
earlier, so the manifold terminates before the first bounce can occur.

The signature criterion engages BKL dynamics in an unusually direct way,
because a Kasner metric is already an anisotropic deformation of exactly the
kind treated in Sec.~\ref{subsec:stress}. A squashed Kasner element carries a
small spatial eigenvalue by construction, so it sits nearer to
positive-definiteness failure than an isotropic element of the same volume.
The criterion is therefore not neutral with respect to anisotropy: it removes
the most squashed configurations first, and in that specific sense acts as an
isotropizer on the surviving ensemble.

BKL chaos relies on crests. Individual Kasner crests temporarily drive local
regions to large anisotropic deformation on scale $q$ at radii $r > \rBQ$.
When a localized crest reaches the threshold $\Tqb(\alpha)$, the elements in
it are no longer Lorentzian, the crest is absent from the manifold, and the
energy that would have been carried forward into the next Kasner epoch is not
carried anywhere. The crest becomes a transient, localized $\BQ$ boundary.

Two things should be distinguished here. What the criterion removes is the
chaos, not the anisotropy. An oscillation is carried forward by its crests and
troughs together, so excising the crests does not leave a clipped wave that
continues with reduced amplitude: the troughs alone cannot propagate it,
because the elements through which it would travel are no longer in the
manifold. Connected support is severed along the very direction the succession
would advance. The sequence of epochs therefore ends at the first truncated
crest rather than being held in check by any restoring mechanism, and the
anisotropy already present is left as it stands, since nothing in the
criterion restores isotropy. And the truncation is geometric rather than
dissipative. Inside a black hole horizon $\partial_r$ is not generally a
Killing vector, so it generates no conserved charge; at a truncated crest no
continued local geometry exists on which Killing symmetries or conserved
charges could be extended. The exterior ADM mass and spin remain fixed by the
admissible exterior geometry. The next manifold element is simply not there.

\subsection{Weak and Strong Cosmic Censorship} \label{subsec:cosmic-censorship}

The $\BQ$ framework provides a physical foundation for both weak and strong cosmic censorship. Mass inflation destabilizes the inner Cauchy horizon before it can function as a gateway to the regular Kerr extension or CTC sector, supporting strong censorship by eliminating the non-unique extension. Concurrently, $\BQ$ supplies the quantum-geometric termination of the remaining high-stress region, supporting weak censorship by preventing naked singular exposure.

\section{Conclusion} \label{sec:conclusion}

This paper proposes a resolution of the black hole singularity problem by
asserting that a local, algebraic, coordinate-invariant admissibility
criterion on the metric terminates the interior geometry. Operating within the
Wheeler--DeWitt framework supplemented by the admissibility and connectivity
conditions of Eqs.~\eqref{eq:admissibility_chain} and
\eqref{eq:connectivity_chain}, the $\BQ$ boundary establishes several
core results:

\begin{itemize} 

	\item \textbf{Feynman Sum Exclusion:} The sum over geometries provides the full geometrical wavefunctional on its admissible domain. Where a local realization of $g_{\mu\nu}$ is not Lorentzian there is no metric element, hence no spacetime, and configurations requiring one have vanishing support ($\Psi = 0$). The vacuum fluctuation amplitude $\Delta_g = \sqrt{8/\pi}\,\lP/q$ grows without bound as the scale falls, which sets a floor $q_{\min}$ on the size of a spacetime element. Where curvature adds a deterministic deformation to that fluctuation, a fraction of the elements ceases to exist, and there is a first surface at which the surviving Lorentzian elements can no longer sustain a connected differentiable support.
	\item \textbf{A Spherical Reference Radius:} For the vacuum Schwarzschild interior the tidal Weyl stress alone reaches threshold at $\rBQ \simeq 1.4\,(r_g\lP^2)^{1/3}$, or $1.0\times10^{-22}(M/M_\odot)^{1/3}$~m, thirteen orders of magnitude above $\lP$ and at finite curvature.
	\item \textbf{A Cutoff for Mass Inflation:} For realistic Kerr black holes the operative stress is not tidal but the trace-free Ricci stress of the counter-streaming null fluxes, which is larger by the ratio $2r_{-}/d$ of the inner-horizon radius to the thickness of the inflating layer. That stress acts on a single axis and extinguishes the radial direction, and because it grows as the layer thins while $d$ is floored at one spatial quantum, no configuration of the streams evades it. The internal mass parameter is capped at $\nBQ$ of Eq.~\eqref{eq:nmaxcases}, terminating the manifold at the inner horizon and placing the classical ring, the Cauchy horizon, and the closed timelike curves of the extended vacuum solution outside the physical spacetime.
	\item \textbf{An Accretion-Rate Dependence:} Because the layer thickness is set by the accretion rate and is floored at one spatial quantum, $\nBQ$ depends on $\mu$, ranging over $\ln\nBQ \approx 88$ to $157$ for a $10\,M_\odot$ hole: the cap is not a property of the hole alone but of its accretion history.
	\item \textbf{Physical Action of the Truncation:} The Lorentzian evaluation of the Gibbons--Hawking--York boundary term over the terminal spacelike slice yields a finite macroscopic action ($S_{GHY}^{\rm qb} \approx \frac{3}{2}\nBQ Mc^2\,\Delta t$).
	\item \textbf{Capping BKL Oscillations:} As anisotropic Kasner crests reach the deformation threshold, their elements leave the manifold. Local continuous symmetries and their corresponding conservation laws have no continued geometry on which to be extended, allowing these transient $\BQ$ boundaries to cap structural divergence without violating energy or momentum conservation.
\end{itemize}

Alternative approaches, such as loop quantum cosmology \citep{bojowald2001}, spinor-torsion bounces, and Planck stars \citep{rovelli2014}, seek to resolve the singularity by injecting novel trans-Planckian physics. By replacing the continuous manifold with discrete area spectra, they kinematically force a quantum bounce. However, this reliance on additional structural assumptions introduces profound observational vulnerabilities. Should future high-energy astrophysics or advanced interferometry confirm that spacetime remains a continuous fabric down to the Planck scale, frameworks requiring a pixelated or discrete spacetime lattice would be substantially disfavored. Furthermore, while these models demonstrate that a bounce is mathematically possible in idealized, perfectly homogeneous settings, they do not show that a coherent bounce dynamically occurs in a causally isolated, BKL-scrambled black hole interior. Such an observation would bear differently on the present framework, because $q_{\min}$ is not a lattice spacing. It is a floor on the size of a coherent metric element rather than a distance between points, and the vacuum is not smooth at that scale: at $\Delta_g$ of order unity a substantial fraction of elements fail, so flat space is threaded with transient quantum topological holes even where connected differentiable support is easily maintained. What the framework does not contain is a persistent structure. The holes are stochastic and short lived, with no fixed spacing, no preferred direction, and nothing that recurs from one passage to the next, so the signatures by which discreteness is usually sought; vacuum birefringence, energy-dependent arrival times, a stationary noise floor at a fixed scale, have no source here, and a null result in those searches does not disfavor it. A randomly excised vacuum may nonetheless leave a diffusive imprint on propagation over long baselines, of the $\sqrt{\lP L}$ character noted in Sec.~\ref{subsec:vacuum-amplitude}. That has not been computed here, but it suggests the framework may be testable in flat space as well as through the interior prediction of Eq.~\eqref{eq:nmaxcases}. In the present framework, no additional terms in the Wheeler--DeWitt Hamiltonian rescue the geometry at $\BQ$: where the metric is not Lorentzian, every continuous metric-based operator loses its argument simultaneously.

We have shown that classical general relativity combined with standard quantum
mechanics naturally limits its own collapse within the geometrical domain. What
replaces the singularity is not a crushing of matter to a point but the end of
the interior's time: because $r$ is timelike within the horizons, $\BQ$ is a
spacelike surface, and an infalling body reaches it whole and at once, at the
last moment the manifold contains. The
criterion itself, the numerical determination of $q_{\min}$, and the
thresholds used here are developed in \citep{shaya2026signature}. The extension of the $\BQ$ boundary to cosmology and to the full configuration
space of geometries is examined in a companion paper \citep{shaya2026_mwi}.

\appendix

\section{Finite-scale response coefficients from the normal-coordinate
deformation}
\label{app:CR}

The admissibility criterion tests the signature of the metric over a region of
extent $q$, so the object it responds to is the deformation of the metric
across that region rather than any single curvature scalar. In Riemann normal
coordinates about a point,
\begin{equation}
	g_{\mu\nu}(x) \;=\; \eta_{\mu\nu}
	\;-\; \tfrac13 R_{\mu\alpha\nu\beta}\,x^{\alpha}x^{\beta}
	\;+\; \mathcal{O}(x^3),
	\label{eq:normalcoord}
\end{equation}
and for a spatial displacement $x^i = q\,n^i$ with $n^i$ a unit vector the
radial entry is depressed by $\delta g_{zz} = -\tfrac13 q^2
R_{z i z j}n^i n^j$. Extinction of that direction requires $\delta g_{zz}=-1$.
We evaluate $R_{\mu\alpha\nu\beta}$ in a local orthonormal frame with
signature $(-,+,+,+)$, $z$ along the streaming axis and $x,y$ transverse, and write
\begin{equation}
	\lambda \;\equiv\; \frac{4\pi G\rho}{c^{2}} .
\end{equation}

\subsection{Counter-streaming null fluxes}

Opposed null fluxes of equal intensity along $\pm z$ have
$T_{\mu\nu} = \rho c^2\operatorname{diag}(1,0,0,1)$ in the frame in which
their momenta cancel. The trace vanishes identically, so $R=0$ and
$R_{\mu\nu} = (8\pi G/c^{4})T_{\mu\nu} = 2\lambda
\operatorname{diag}(1,0,0,1)$. With the Weyl part set to zero, as in the Ricci
response of Eq.~\eqref{eq:ricciresponse}, the Riemann tensor is reconstructed
from
\begin{equation}
	\begin{split}
	R_{\mu\alpha\nu\beta} = \tfrac12\bigl(
	&\eta_{\mu\nu}R_{\alpha\beta} + \eta_{\alpha\beta}R_{\mu\nu}\\
	&- \eta_{\mu\beta}R_{\alpha\nu} - \eta_{\alpha\nu}R_{\mu\beta}\bigr),
	\end{split}
	\label{eq:riemannfromricci}
\end{equation}
which gives
\begin{equation}
	\begin{split}
	&R_{zxzx}=R_{zyzy}=\lambda,\qquad R_{ztzt}=0,\\
	&R_{xyxy}=0,\qquad R_{txtx}=R_{tyty}=\lambda .
	\end{split}
	\label{eq:riemannnull}
\end{equation}
As a check, $R_{zz} = -R_{ztzt}+R_{zxzx}+R_{zyzy} = 2\lambda$ and
$R_{xx} = -R_{xtxt}+R_{xyxy}+R_{xzxz} = -\lambda+0+\lambda = 0$, as required.

\subsection{The threshold and the response coefficient}

From Eq.~\eqref{eq:riemannnull}, a spatial displacement in the plane
transverse to the streaming axis gives
\begin{equation}
	\delta g_{zz} = -\tfrac13\lambda q^{2}\left[(n^{x})^{2}+(n^{y})^{2}\right],
\end{equation}
so extinction of that direction requires
\begin{equation}
	\begin{split}
	\lambda q^{2} &= 3 \quad\text{(transverse displacement)},\\
	\lambda q^{2} &= 4.5 \quad\text{(direction averaged)},
	\end{split}
	\label{eq:lambdathreshold}
\end{equation}
that is, $8\pi G\rho\,q^{2}/c^{2} = 6$ to $9$. The tidal block
$R_{0i0j}=\lambda\operatorname{diag}(1,1,0)$ enters $g_{tt}$ alone and drives
it further from zero, so it cannot produce signature failure.

Equation~\eqref{eq:ricciresponse} is written against $S_{\mu\nu}$, and for
traceless null stress $S_{zz} = R_{zz} = 2\lambda$, so
Eq.~\eqref{eq:lambdathreshold} corresponds to $S_{zz}q^{2} = 6$ to $9$ and
\begin{equation}
	C_R \;=\; \tfrac16 \text{ to } \tfrac19 \;\simeq\; 0.15 .
\end{equation}

The estimate inherits the truncation of Eq.~\eqref{eq:normalcoord} at second
order, and it is worth asking how large the next term is. The cubic term and
the $\nabla\nabla R$ part of the quartic carry derivatives of curvature, and
with $|\nabla R|/|R| \sim 1/r$ in a black hole interior they are suppressed
relative to the quadratic term by $q/r \sim 10^{-38}$. Only the $R^2$ part of
the quartic term matters,
\begin{equation}
	\delta g^{(4)}_{\mu\nu} = \tfrac{2}{45}
	R_{\mu\alpha\lambda\beta}R^{\lambda}{}_{\gamma\nu\delta}\,
	x^{\alpha}x^{\beta}x^{\gamma}x^{\delta} .
	\label{eq:quartic}
\end{equation}
Evaluated on Eq.~\eqref{eq:riemannnull} and averaged over the transverse
plane, its ratio to the quadratic term is $0.13$ at $\lambda q^2 = 1$, $0.27$
at $2$, and $0.40$ at the threshold $\lambda q^2 = 3$. The correction is
therefore substantial but ordered, and $C_R$ should be read as carrying an
uncertainty of roughly a factor of two rather than as a determination.

\section{The Weyl response coefficient}
\label{app:CW}

The same construction fixes $C_W$. For the vacuum Schwarzschild interior the
Ricci tensor vanishes and the Riemann tensor is pure Weyl. In a local
orthonormal frame adapted to the static observer, with
$\hat r$ radial and $\hat\theta,\hat\varphi$ transverse, and with
\begin{equation}
	\mu \;\equiv\; \frac{r_g}{r^{3}} ,
\end{equation}
the independent components are
\begin{equation}
	\begin{split}
	&R_{\hat t\hat r\hat t\hat r} = -2\mu,\qquad
	R_{\hat t\hat\theta\hat t\hat\theta}
	= R_{\hat t\hat\varphi\hat t\hat\varphi} = \mu,\\
	&R_{\hat r\hat\theta\hat r\hat\theta}
	= R_{\hat r\hat\varphi\hat r\hat\varphi} = -\mu,\qquad
	R_{\hat\theta\hat\varphi\hat\theta\hat\varphi} = 2\mu ,
	\end{split}
	\label{eq:riemannschw}
\end{equation}
with $R_{\hat\mu\hat\nu}=0$ on every contraction, as required. The electric
part is $E_{ij} = R_{\hat 0 i \hat 0 j} = \mu\operatorname{diag}(-2,1,1)$,
reproducing Eq.~\eqref{eq:schwT}.

Inserting Eq.~\eqref{eq:riemannschw} into the normal-coordinate deformation
\eqref{eq:normalcoord}, with the displacement averaged over the plane
perpendicular to the direction being tested as in
Appendix~\ref{app:CR}, gives for a transverse direction
$\langle n^{k}n^{l}\rangle = \tfrac12\operatorname{diag}(1,0,1)$ and
\begin{equation}
	\delta g_{\hat\theta\hat\theta}
	= -\tfrac13 q^{2}\cdot\tfrac12\left(
	R_{\hat\theta\hat r\hat\theta\hat r}
	+ R_{\hat\theta\hat\varphi\hat\theta\hat\varphi}\right)
	= -\frac{\mu q^{2}}{6},
\end{equation}
while for the radial direction
$\langle n^{k}n^{l}\rangle = \tfrac12\operatorname{diag}(0,1,1)$ gives
\begin{equation}
	\delta g_{\hat r\hat r}
	= -\tfrac13 q^{2}\cdot\tfrac12\left(
	R_{\hat r\hat\theta\hat r\hat\theta}
	+ R_{\hat r\hat\varphi\hat r\hat\varphi}\right)
	= +\frac{\mu q^{2}}{3} .
\end{equation}
The deformation is therefore $\mu q^{2}\operatorname{diag}(2,-1,-1)/6$: the
normal-coordinate expansion reproduces the eigenvalue pattern assumed in
Eq.~\eqref{eq:schwT} rather than merely being consistent with it, with the
radial direction stretched and the two transverse directions compressed.
Comparing with $H_{ij}$ of Eq.~\eqref{eq:schwarzschildH} and using
$\mathcal{T} = \mu q^{2}$,
\begin{equation}
	C_W \;=\; \tfrac16 ,
\end{equation}
or $C_W = \tfrac19$ if the displacement is averaged over the full sphere,
$\langle n^{k}n^{l}\rangle = \tfrac13\delta^{kl}$, which leaves the
$2:-1:-1$ pattern unchanged and rescales only the magnitude. We adopt
$C_W = 0.15$.

That $C_W$ and $C_R$ fall in the same range is not accidental. Both are
$\tfrac13$ times a transverse average of two curvature components, and the
factor of two between the transverse-plane and full-sphere conventions is
common to them. We take $C_W = C_R = 0.15$ throughout, retaining the symbols so
that the scaling of each result with its response coefficient stays visible.

The higher-order behavior differs from that of Appendix~\ref{app:CR} in a way
worth identifying. Evaluating Eq.~\eqref{eq:quartic} on
Eq.~\eqref{eq:riemannschw} and averaging over the transverse plane, the
quartic term reaches the size of the quadratic one near $\mu q^2 \sim 3$ and
carries the opposite sign. This is not a repulsive correction, which the
Einstein equations forbid for a positive mass. What a positive quartic term
registers is the closing of the exponential map, the reconvergence of
geodesics that occurs in a positively curved sector; in normal coordinates the
vanishing of a metric component marks a conjugate point, where the coordinates
themselves cease to be valid. For a maximally symmetric space the expansion
resums exactly to $(\sin u/u)^2$, whose zero at $u=\pi$ is such a conjugate
point. Schwarzschild is not maximally symmetric and the resummation is not
that function, but the mechanism is the same: a vanishing of the truncated
$g_{\hat\theta\hat\theta}$ in this sector would be a caustic of the coordinate
system rather than a loss of Lorentzian signature.

The tidal case therefore never reaches a degenerate metric, and does not need
to. What the Einstein equations fix, and what no higher-order term alters, is
the sign and pattern of the deformation: radial stretching with transverse
compression, $\operatorname{diag}(2,-1,-1)$, which is what the field of a
positive mass produces. $C_W = 1/6$ sets the normalization of that pattern and
should be read as an order-of-magnitude value. The threshold itself is not
taken from it: $\Tqb = 0.35$ in Sec.~\ref{subsec:schwarzschild-radius} comes
from the fluctuation ensemble at a background deformation of
$C_W\Tqb \approx 0.05$, where truncation at second order is reliable and the
quartic term of Eq.~\eqref{eq:quartic} is smaller again by that factor.

\section*{Acknowledgments}

The author would like to thank the late J.~A.~Wheeler, who, many years ago, revealed to me, among many things, the overarching significance of $H\Psi = 0$.
The author acknowledges the computational resources provided by the
University of Maryland. Generative artificial intelligence played a
significant role in developing the mathematical formulation and the
numerical code used to obtain the results.  The author independently
verified the scientific arguments, calculations, and conclusions.  Various models were usefully engaged to provide competing numerical codes and to make critical evaluations.  The drafts were circulated periodically between Anthropic models Opus 5 and Fable 5.1, OpenAi 5.6 Sol  and Tera, and Gemini 3.1 Pro and 3.8 Flash.

\bibliographystyle{apsrev4-2}
\bibliography{high_densities}

\end{document}